\documentclass[11pt,a4paper]{article} 
\usepackage{epsfig} 

\author{Tom Weidig, University of Durham, UK\\ email: tom.weidig@physics.org}
\title{Quantum Mass Correction of Solitons\\in (1+1)D via Numerical Methods}

\begin{document}

\maketitle

\abstract{We show how to calculate the quantum mass correction to
(1+1)D solitonic field theories using numerical methods. This is
essential if we want to find the corrections to non-integrable models.
We start with  a review of the standard derivation of the first order
quantum correction.  Then, we re-derive a trace formula which allows
us to compute the mass correction mode by mode.  Specifically, we are
interested in  the extent to which the lowest modes from both, the
soliton and the vacuum, sectors give the leading contribution. We apply the
 technique to both the Sine-Gordon and the $\phi^4$-kink model. Then,
we compute all the modes numerically and hence the first order quantum
contribution to the mass of the Sine-Gordon and $\phi^4$ soliton.}

\section{Introduction}

We focus on the computation of the mass correction  of
(1+1)D solitonic theories theoretically and numerically. We follow
Rajaraman's semi-classical quantisation procedure; another option is via path
integrals -- see Dashen et al. \cite{dashen}.  Specifically, we are
interested in the use of the normal modes to compute the mass correction and
the extent to which the lowest modes from both, the soliton and the vacuum,
sectors give the leading contribution. We start
with  a review of the derivation of the first order quantum correction.
 Then, we derive the trace formula from first
principles.\footnote{this is not done in \cite{trace:cahill}.}  We use this
formula to re-compute the mass correction and show that the lowest modes are
the most important ones. Then, we calculate the lowest modes numerically and
hence the leading contribution to the mass correction. 

\section{Mass Quantum Correction: General Idea}

We start out with a general lagrangian for a (1+1)-dimensional field theory
\begin{equation}
{\cal L}=\frac{1}{2}\dot{\phi}^2-\frac{1}{2}{\phi'}^2-V(\phi)
\end{equation}
for the scalar field $\phi(t,x)$ and with the potential $V$ being positive. The
time-independent Euler-Lagrange equation leads to 
\begin{equation}\label{static}
-\phi''+\frac{d V}{d \phi}=0.
\end{equation}
We quantise around the minimal-energy static solution
$\phi_{st}(x)$ satisfying (\ref{static}), which could be the vacuum or the
minimal-energy solution in a non-zero topological sector. The semi-classical
expansion states that the quantum field $\hat{\phi}(t,x)$ is the classical
static field $\phi_{st}(x)$ plus a quantum correction field 
$\hat{\epsilon}(t,x)$,
\begin{equation}\label{semi-classical}
\hat{\phi}(t,x)=\phi_{st}(x)+\hat{\epsilon}(t,x)
\end{equation}
where~$\hat{}$~reminds us to treat the function as a quantum object, an
operator satisfying, possibly non-commuting, commutation relations with
other operators. We have to substitute (\ref{semi-classical}) into the
hamiltonian 
\begin{equation}\label{qenergy}
H\left(\phi\right)=\int_{-\infty}^{\infty}dx
\left(\frac{1}{2}\dot{\phi}^2+\frac{1}{2}{\phi'}^2+V(\phi)\right)
\end{equation}
and obtain, in orders of $\epsilon$, 
\begin{equation}
\hat{{\cal H}}(x)=\underbrace{\frac{1}{2}\phi_{st}'^2}_{\mbox{Classical}}+
\hat{\epsilon}\underbrace{\left(-\phi_{st}''+\left.\frac{d V}{d \phi}
\right|_{\phi_{st}}\right)}_{\mbox{$=0$ (\ref{static})}}
+\underbrace{\frac{1}{2}\hat{\pi}^2\ +\frac{1}{2}\hat{\epsilon}\left(
-\frac{d^2}{dx^2}+\left. \frac{d^2 V}{d
\phi^2}\right|_{\phi_{st}}\right)\hat{\epsilon}
+O(\hat{\epsilon}^3}_{\mbox{Quantum}}) 
\end{equation}
where $\hat{\pi}$ is the conjugate momentum of $\hat{\epsilon}$. We 
have used integration by parts, set the boundary terms to zero and
Taylor expanded the potential term in powers of $\epsilon$. We  split our
hamiltonian into three parts,
\begin{equation}\label{hamilton:split}
\hat{H}= H_{Classical}+\hat{H}_{Quantum}+\hat{H}_{HigherOrder},
\end{equation}
where the classical mass/energy can be found by substituting $\phi_{st}$ into
(\ref{qenergy}). We concentrate on the lowest order quantum energy
$\hat{H}_{Quantum}$. This is justified in the framework of a perturbation
theory where higher order terms are neglected due to their smallness in
comparison to the lowest term. This approximation is justified if the
potential is roughly harmonic around the static solution and $\epsilon^n$ terms
depend on the coupling constant $\lambda$ in the form $\lambda^{n}$, for
example. If these conditions are not fulfilled or a better accuracy is
wanted, one can resort to well-known perturbation techniques of standard
quantum mechanics.  However, let us emphasise that this would considerably
complicate our task. Therefore, Barnes et al.'s attempts are already
significantly hampered for those multi-skyrmions whose potential does not
have a steep valley i.e.\ are not suitable for a harmonic oscillator
approximation. This is the case for $B=2$, for example.\footnote{private
communication from Barnes} 

Our quantum hamiltonian, in the harmonic approximation, has the form
\begin{equation}\label{qhamilton}
2\hat{{\cal H}}(t,x)=\hat{\pi}^2(t,x) +\hat{\epsilon}(t,x)
\underbrace{\left( -\frac{d^2}{dx^2} + \left.
\frac{d^2V}{d\phi^2}\right|_{\phi_{st}} \right) }_{{\bf
A^2}}\hat{\epsilon}(t,x)
\end{equation}
where $A^2$ is an operator. If $A^2$ acts as a number, the hamiltonian
has the form of an harmonic oscillator and we know the quantisation procedure.
In effect, we have to solve the eigenvalue equation
\begin{equation}
{\bf A^2}\hat{\epsilon}(t,x)=\omega^2\hat{\epsilon}(t,x)
\end{equation}
and this is equivalent to the time-independent Schr\"odinger equation. We have
to decompose the quantum field $\hat{\epsilon}$ in terms of a complete set of
real and orthonormal eigenfunctions. Therefore,
\begin{equation}\label{qdecomposition}
\hat{\epsilon}(t,x)=\sum_n^{\infty}~\hat{q}_n(t)\hat{\eta}_n(x)
\end{equation}
with 
\begin{eqnarray}\label{qrelations}
\sum_{n}\hat{\eta}_n(x)\hat{\eta}_n(y)&=&\delta(x-y)\nonumber\\ 
\int dx\left[\hat{\eta}_n(x)\hat{\eta}_m(x)\right]&=&\delta_{n,m}
\end{eqnarray}
and
\begin{equation}
{\bf A^2}\hat{\eta}_i(x)=\omega^2_i\hat{\eta}_i(x).
\end{equation}
We substitute the decomposition (\ref{qdecomposition}) into
(\ref{qhamilton}), integrate over $x$ and use the constraints 
(\ref{qrelations}) on the eigenfunctions. We are left with
\begin{equation}
2\hat{{\cal H}}(t)=\sum_n\left[\hat{p}_n^2(t)+\omega^2_n\hat{q}_n^2(t)\right]
\end{equation}
where $\hat{p}_n$ is the conjugate momentum  of $\hat{q}_n$. The hamiltonian
is an  infinite sum of harmonic oscillators of frequency $\omega_n$. We have
reduced our quantum field theory with an infinite number of degrees of
freedom to a pseudo particle quantum mechanics with  an infinite number of
harmonic oscillators. We are now able to use standard particle quantum
mechanics. Using the operator method with Heisenberg's commutation relation
\begin{equation}
\left[\hat{p}_n,\hat{q}_n\right]=i\hbar,
\end{equation}
we get
\begin{equation}
H=\hbar\sum_n\left(\alpha_n+\frac{1}{2}\right)\omega_n
\end{equation}
where $\alpha_n$ is the $\alpha$th energy level of the nth oscillator.
Naively speaking, we have all the information we need to calculate the mass
correction: the classical mass and the quantum correction to first order
i.e.\ the quantum hamiltonian. Unfortunately, if we were to calculate the
quantum mass in  a specific model, we would quickly realise that the mass is
divergent; the infinite number of oscillators, an inherent feature of any
quantum field theory, being the cause of this divergent result. In the next 
section, we describe, using the $\phi^4$ model as our example, how to extract
a meaningful quantum properties from our naive expression. 

\section{Mass Quantum Correction: Derivation}

We follow Rajaraman's procedure. For a very detailed and lucid description,
we refer to Rajaraman's book \cite[section 5.3]{raj}. We have re-done all the
calculations by hand and with the use of Maple. Further we have expanded on
the discussion of some parts, given more details and  corrected some 
typographical errors.\footnote{and probably introduced others.}

The hamiltonian of the $\phi^4$ kink model is 
\begin{equation}
2H=\int dx\left( \dot{\phi}^2+\phi'^2-m^2\phi^2
+\frac{\lambda}{2}\phi^4+\frac{m^4}{2\lambda}\right)
\end{equation}
where $m$ is the mass of the field $\phi$ and $\lambda$ the self-coupling
constant. In topological charge sector zero, the minimal-energy solution
i.e.\ the vacuum is 
\begin{equation}
\phi_{st}(x)=\pm\frac{m}{\sqrt{\lambda}}
\end{equation}
and, in topological charge sector one, the minimal-energy solution i.e.\ the
1-kink is 
\begin{equation}
\phi_{st}(x)=\pm\frac{m}{\sqrt{\lambda}}\tanh
\left[\frac{m(x-a)}{\sqrt{2}}\right]
\end{equation}
where $a$ indicates a translational invariance (this will lead to a zero
mode). We quantise around the solutions with the positive sign in front which
is the soliton --  compared with the negative sign in front which is the
anti-soliton.

The corresponding eigenvalue equation for the vacuum is
\begin{equation}
\left[-\frac{d^2}{dx^2}+2m^2\right]\hat{\eta}_i(x)
=\omega_{V,i}^2~\hat{\eta}_i(x)~.
\end{equation}
The eigenfunctions and eigenvalues are
\begin{equation}
\hat{\eta}_k(x)=\exp(ikx)
\end{equation}
and
\begin{equation}
\omega_{V,k}^2=k^2+2m^2.
\end{equation}
We use periodic boundary conditions in a box of length $L$ and 
\begin{equation}\label{pbc:k}
k_nL=2\pi n
\end{equation}
where $n$ is an integer.
The continuum limit is reached by taking $L$ to infinity and any discrete sum
over $k_n$, or simply $n$, turns into an integral over $k$ of the form
\begin{equation}\label{int:k}
\sum_{n}\longrightarrow \int dn=\frac{L}{2\pi}\int dk
\end{equation}
using the constraint (\ref{pbc:k}) on $k_n$.

The corresponding eigenvalue equation for the 1-kink is less trivial
\begin{equation}
\left[-\frac{1}{2}\frac{d^2}{dz^2}+(3\mbox{tanh}^2z-1)\right]\hat{\eta}_i(z)
=\frac{\omega_{K,i}^2}{m^2}~\hat{\eta}_i(z),
\end{equation}
where $z\equiv mx/\sqrt{2}$ for convenience. In fact, it belongs to a class of
special Schr\"odinger equations; the Sine-Gordon model being another
example.\footnote{private communication from Jackiw and see \cite[pages
683-684]{jackiw}}\label{jack} There are two discrete modes; one zero mode
\begin{equation}\label{phi:modes}
\hat{\eta}_0(z)=\cosh^{-2}z~\mbox{~with~}~\omega_{K,0}^2=0  
\end{equation}
and a second discrete mode
\begin{equation}
\hat{\eta}_1(z)=\sinh z~\cosh^{-2}z~\mbox{~with~}~\omega_{K,0}^2
=\frac{3}{2}m^2.
\end{equation}
The continuous eigenmodes, which we label with $q$, are 
\footnote{the eigenmodes are Jacobi polynomials in $\tanh z$. \cite{jackiw}} 
\begin{equation}
\hat{\eta}_q(z)=e^{iqz}\left(3\tanh^2z-1-q^2-3iq \tanh z\right)
\end{equation}
with
\begin{equation}
\omega_{K,q}^2=m^2\left(\frac{q^2}{2}+2\right).
\end{equation}
Imposing periodic boundary conditions becomes more tricky. For
$z\rightarrow\pm\infty$, we can use the asymptotic form of $\hat{\eta}_q(z)$
\begin{eqnarray}
\hat{\eta}_q(z)&\longrightarrow& e^{iqz} \left(2-q^2\mp3iq\right)\nonumber\\
&\longrightarrow&\sqrt{(q^2+4)^2(q^2+1)}
\exp\left[i(qz\pm\frac{1}{2}\delta(q))\right]
\end{eqnarray}
where $\delta(q)$ is the phase shift of the scattering states from the
viewpoint of the Schr\"odinger equation;
\begin{equation}\label{delta:q}
\delta(q)=-2~\mbox{arctan}\left[\frac{3q}{2-q^2}\right].
\end{equation}
The condition imposed by the periodic boundary is
\begin{equation}\label{pbc:q}
q_n\left(\frac{mL}{\sqrt{2}}\right)+\delta(q_n)=2\pi n
\end{equation}
and the sum over $n$ becomes an integral over $q$ in the limit where $L$ goes
to infinity:
\begin{equation}
\sum_n\longrightarrow\int dn =\frac{1}{2\pi}\int dq
\left(\frac{mL}{\sqrt{2}}+\frac{\partial}{\partial q}\left[\delta(q)\right]
\right)\nonumber
\end{equation}
using the constraint (\ref{pbc:q}) on $n$. Let us briefly note that the zero
mode, which we have found, is nothing else but the manifestation that the
1-kink solution is translationally invariant. We can imagine the
infinite-dimensional potential space at its minimal-energy location having a
1-dimensional valley along which we can move our solution by varying $a$
without changing the energy of the system. This zero mode certainly needs to
worry us, for our harmonic oscillator approximation assumes steep valleys in
all dimensions. However, this is only problematic if the zero mode is coupled
with another mode and this does not happen in the computation of the mass
correction up to first order. A proper treatment of zero modes is done with
collective coordinates: see Rajaraman \cite[chapter 8]{raj}. 

We have all the necessary information on the eigenvalues and should naively
be able to compute the mass of the 1-kink up to first order quantum
corrections. Using (\ref{hamilton:split}) and the eigenvalues of the 1-kink
solution, we get an expression for the energy
\begin{equation}
E_K=\underbrace{\underbrace{\frac{2\sqrt{2}m^3}{3\lambda}}_{Classical}
+\underbrace{\hbar m\frac{\sqrt{3}}{2\sqrt{2}}}_{Discrete}}_{Finite}
+\underbrace{\frac{1}{2}\hbar m\underbrace{\sum_n
\sqrt{\frac{1}{2}q_n^2+2}}_{Continuous}}_{Divergent}
\end{equation}
which includes the finite classical energy, no contribution from the zero
mode due to its zero frequency, a finite contribution from the second
discrete mode and a sum over the continuous modes.  Unfortunately, if we
were to perform the integral over $q$ using (\ref{pbc:q}), we would find it
to be divergent. This clearly shows that our naive treatment of quantum field
theory is inadequate. To have a finite answer, there are two modifications we
have to make. 

\subsection{Energy level difference}
Let us write out the expression for the vacuum energy up to first order
quantum corrections
\begin{equation}
E_V=\frac{1}{2}\hbar\sum_n\sqrt{k_n^2+2m^2}.
\end{equation}
Using (\ref{int:k}), we get 
\begin{equation}
E_V=\frac{\hbar L}{4\pi}\int_{-\infty}^{\infty}dk\sqrt{k^2+2m^2}
\end{equation}
which is a quadratically divergent integral. Thus, 
even the leading quantum contribution to the classical vacuum is not
finite. However, we can follow the example of newtonian gravity which defines
potential energy as the difference between two states. It makes physical
sense to define our naive vacuum energy, even though it is infinite, as the
lowest of possible energy states in the theory i.e.\ to put it equal to zero
and hence to subtract it from our naive calculation of the 1-kink energy. We
get
\begin{equation}
E_K-E_V=\tilde{E}_K= \underbrace{\frac{2\sqrt{2}m^3}{3\lambda} +\hbar
m\frac{\sqrt{3}}{2\sqrt{2}} }_{E_{finite}}
+\frac{1}{2}\hbar\sum_n\left(m\sqrt{\frac{1}{2}q_n^2+2}-\sqrt{k_n^2+2m^2}
\right)
\end{equation}
and we label all finite terms collectively $E_{finite}$. We go to the
continuum limit and perform the integral over $k$. Therefore, we re-express
$q_n$ in terms of $k_n$
\begin{equation}
q_n=\frac{\sqrt{2}}{m}\left(k_n-\frac{\delta(q_n)}{L}\right)
\end{equation}
using the boundary conditions (\ref{pbc:q}) and (\ref{pbc:k}). The expression
in the sum takes the form
\begin{eqnarray}
m\sqrt{\frac{1}{2}q_n^2+2}-\sqrt{k_n^2+2m^2}&=&
\sqrt{\left(k_n-\frac{\delta(k_n)}{L}\right)^2+2m^2}-\sqrt{k_n^2+2m^2}\nonumber\\
&=&-\frac{k_n\delta(k_n)}{L\sqrt{k_n^2+2m^2}}+O(1/L^2)\nonumber\\
&=&\frac{2}{L}~\mbox{arctan}\left(\frac{3m}{\sqrt{2}}\frac{k_n}{m^2-k_n^2}\right)
\frac{k_n}{\sqrt{k_n^2+2m^2}}+O(1/L^2)
\end{eqnarray}
where we have Taylor expanded the first line, used expression (\ref{delta:q})
and Taylor expanded it. Both Taylor expansions are in $\frac{1}{L}$ and make
sense, for we take the box size $L$ to infinity later. Using (\ref{int:k}), the
expression for the energy becomes
\begin{equation}
\tilde{E}_K=E_{finite}+\frac{\hbar}{2\pi} \int
dk~\underbrace{\mbox{arctan}\left( \frac{3m}{\sqrt{2}}
\frac{k}{m^2-k^2} \right) }_{du/dk}
\underbrace{\frac{k}{\sqrt{k^2+2m^2}}}_{v}~+~O(1/L)~.
\end{equation}
The dependence on the box size goes away for $L\rightarrow\infty$
and we are
allowed to neglect the $O(1/L)$ and higher terms. As our notation indicates,
we perform integration by parts. The boundary term has the form
\begin{equation}
\frac{\hbar}{2\pi}~\lim_{\alpha\rightarrow\infty}\left[\mbox{arctan}\left(
\frac{3m}{\sqrt{2}} \frac{k}{m^2-k^2}
\right)~\sqrt{k^2+2m^2}\right]_{k=\alpha}^{k=-\alpha}.
\end{equation}
This limit is ill-defined for trivial substitution of $\alpha=\infty$: the 
arctan function gives us 0 and the polynomial function $\infty$. Therefore,
we use l'H\^optial's rule and obtain a finite answer
\begin{equation}
\hbar m~\frac{3}{\pi\sqrt{2}}
\end{equation}
which we include in our $E_{finite}$. The integral obtained by
integration by parts has the form
\begin{equation}
\frac{3\sqrt{2}m}{2\pi}\int dk~\frac{k^2+m^2}{\sqrt{k^2+2m^2}~(2k^2+m^2)}.
\end{equation}
We put a cut-off $\Lambda$ on the $k$ limits and change to the variable
$p\equiv k/m$. We get
\begin{equation}\label{div}
\lim_{\Lambda\rightarrow\infty}\int_{\frac{\Lambda}{m}}^{-\frac{\Lambda}{m}}
\frac{p^2+1}{\sqrt{p^2+2}~(2p^2+1)}
\end{equation}
If we perform this integral, we still find a logarithmic divergence plus a 
finite contribution. We need to cancel the divergence with another term.  We 
need to look closer at the infinities produced by the infinite degrees of 
freedom of a field theory.

\subsection{Normal-Ordering and Counter-terms}

We have to normal-order the hamiltonian and introduce
counter-terms. We do not give a full introduction to
all these more complicated ideas and refer to Ryder \cite{ryder}, for
example, for a detailed introduction. We decompose the field $\phi$ in terms
of a complete set of orthonormal eigenfunctions of the vacuum fluctuations
\begin{equation}
\phi(t,x)=\sum_n\left[ \frac{e^{-i\omega_nt}}{\sqrt{2\omega_n}}\hat{a}_n
\epsilon_n(x)~+~\frac{e^{i\omega_nt}}{\sqrt{2\omega_n}}\hat{a}_n^{\dagger}
\epsilon_n^{\dagger}(x)  \right]
\end{equation} 
where $a$ is the annihilation and $a^{\dagger}$ the creation operator (we
neglect the $\hat{}$ on them). The hamiltonian (\ref{qhamilton}) becomes
\begin{equation}
2\hat{H}=\sum_n\omega_n\left(a_na_n^{\dagger}+a_n^{\dagger}a_n\right)
=\sum_n\omega_n\left(2a_n^{\dagger}a_n+1\right)
\end{equation}
using the orthonormality relations of $\epsilon$ and the commutation relation
between $a_n$ and $a_n^{\dagger}$. The term $a_n^{\dagger}a_n$ is viewed as
the number operator $N_n$ and gives the number of nth oscillators that are
excited. We see that the sum $\sum_{n}1$ is divergent and the common
procedure is to re-define the hamiltonian. We are free to choose the zero of
energy and are allowed to neglect the $1$. Phrased differently, we
normal-order the hamiltonian by writing all annihilation operators to the
right of all creation operators. Thus, we get
\begin{equation}\label{ham:norm}
2:\hat{H}:~=~:\sum_n\omega_n(\underbrace{a_na_n^{\dagger}}_{flip}+a_n^{\dagger}a_n):~=~2\sum_n\omega_na_n^{\dagger}a_n
\end{equation}
where $::$ stands for normal-ordering. The relations between a
normal-ordered and non-ordered product of the fields are
\begin{eqnarray}\label{rel:order}
:\phi^4:~&=&~\phi^4+\alpha\phi^2+\beta\nonumber\\ 
:\phi^2:~&=&~\phi^2+\delta
\end{eqnarray}
where $\alpha$, $\beta$ and $\gamma$ are constants. We write the
normal-ordered hamiltonian as our non-ordered hamiltonian plus two
counter-terms that arise from the relations (\ref{rel:order})
\begin{equation}
:\hat{H}:~=~\hat{H}-\int dx\left(\frac{1}{2}\delta m^2  \phi^2~+~\gamma\right)
\end{equation}
where $\delta m$ is the mass correction to the field and can be evaluated
using the one-loop Feynman diagram. The constant $\gamma$ is not of any
importance, because it will cancel itself out due to its presence in both,
the vacuum and the 1-kink, hamiltonian. The additional term to $E_{finite}$
and the divergent term (\ref{div}) come from subtracting the counter-term
of the vacuum from the counter-term of the 1-kink and we get 
\begin{equation}\label{counter}
E_K^{CT}-E_V^{CT}= \frac{1}{2}\delta m^2\left(\frac{m^2}{\lambda}\right) \int
dx\left[1-~\tanh^2\left(\frac{mx}{\sqrt{2}}\right)\right]
=\frac{\sqrt{2}m}{\lambda}\delta m^2
\end{equation}
We evaluate $\delta m^2$ by using the equivalent expression in $\phi^4$
theory. We refer to Ryder \cite[section 6.4]{ryder} for a detailed
discussion. The standard formula (Ryder: eq 6.95) in perturbation theory for
the $\phi^4$ model is 
\begin{equation}\label{delta:m}
\delta m^2=\frac{1}{2}ig\Delta_{F}(0)
\end{equation}
where $g$ is the coupling and $\Delta_{F}(0)$ the free particle
propagator of a loop diagram i.e $\Delta_{F}(x-x)$. We have to be careful 
when adapting the result to our case. Three modifications to $\phi^4$ 
(Ryder eq 6.65) are necessary:
\begin{itemize}
\item $g/4!=\lambda/4$ and  $g=6\lambda$.
\item The theory should be in (1+1) dimensions.
\item There is only one vacuum. The vacuum eigenvalues are
 $k^2+\tilde{m}^2$ and those of our $\phi^4$ kink theory are
 $k^2+2m^2$. Therefore, we need to change the mass $\tilde{m}^2$ to $2m^2$. 
\end{itemize}
The (1+1) dimensional free particle loop propagator (Ryder: eq 6.14) with 
modified mass $2m^2$ has the form
\begin{equation}
\Delta_F(0)=\frac{\hbar}{(2\pi)^2}\int\frac{d{\bf k}^2}{{\bf k}^2-2m^2}
\end{equation}
where we have a pole at ${\bf k}^2=2m^2$ and the two-vector ${\bf k}$ equals
$(E,-k)$. We evaluate the double integral further
\begin{eqnarray}
\frac{4\pi^2}{\hbar}\Delta_F(0)&=&\int dk\int\frac{dE}{E^2-(k^2+2m^2)}
\nonumber\\ &=&-i\int \frac{dk}{\sqrt{k^2+2m^2}}
\end{eqnarray}
where we have integrated over $E$. We change to the variable $p\equiv k/m$
and put a cut-off $\Lambda$ on $p$. Substituting everything into
(\ref{delta:m}), we get
\begin{equation}
\delta m^2=\frac{3\lambda\hbar}{4\pi}~\lim_{\Lambda\rightarrow\infty}
\int_{-\frac{\Lambda}{m}}^{\frac{\Lambda}{m}} \frac{dp}{\sqrt{p^2+2}}
\end{equation}
which we substitute into the additional term (\ref{counter}).

\subsection{Finite Mass Correction}

Finally, we are able to write the quantum mass of the 1-kink as
\begin{equation}
M=\tilde{E}_K+\tilde{E}^{CT}=E_{finite}+\hbar m~\frac{3\sqrt{2}}
{4\pi}~\lim_{\Lambda\rightarrow\infty}\int_{-\frac{\Lambda}{m}}^{\frac{\Lambda}{m}}
\left[\frac{dp}{\sqrt{p^2+2}}-\frac{2(p^2+1)}{\sqrt{p^2+2}~(2p^2+1)}\right].
\end{equation}
We have done the integral using a cut-off $\Lambda$ with Maple. Both terms
produce the same logarithmic divergent term which cancel each other out.
Taking the cut-off to infinity, we get the final answer for the mass of a
1-kink up to first order quantum corrections 
\begin{equation}\label{full}
M=\left(\frac{2\sqrt{2}}{3}\right)~\frac{m^3}{\lambda}+~\hbar m
\left[\frac{1}{6}\sqrt{\frac{3}{2}}-\frac{3}{\pi\sqrt{2}}\right]
\end{equation}
where we have written $E_{finite}$ out explicitly. The first term is just the
total energy of the classical 1-kink solution. Note that the presence of
$1/\lambda$ indicates the non-perturbative nature of the solution. To zeroth
order in $\lambda$ and first order in $\hbar$, we have the first quantum
correction. It is only valid in the weak-coupling limit. The next term of the
quantum correction would be of order $\lambda\hbar^2$. Rajaraman
\cite[section 5.4-5.6]{raj} gives a detailed interpretation of the result and
also explains why the effect of the counter-terms on the kink solution and
the zero mode are effects of order $\lambda$. 

This concludes our derivation of the quantum mass. In the next section, we
show that it is possible to get a formula for the mass correction which
allows us to quantify the contribution of the different modes and compute the
mass correction numerically by using numerically computed  lowest eigenmodes.

\section{Trace formula: Derivation}

The trace formula has been first published by Cahill et
al. \cite{trace:cahill}, but an explicit derivation has not been given in
their  paper. We derive the formula in this section\footnote{private
communication by Barnes}. We start by writing out the hamiltonian
({\ref{qhamilton}) 
\begin{equation}
2\hat{{\cal H}}(t,x)=\hat{\pi}^2(t,x) +\hat{\epsilon}(t,x){\bf A^2}
\hat{\epsilon}(t,x)
\end{equation}
and the equation of motion of the normal modes is
\begin{equation}
\ddot{\epsilon}(t,x)=-\omega^2\epsilon(t,x)
\end{equation}
with the eigenvalue equation
\begin{equation}
{\bf A^2}~\epsilon_i(x)=\omega_i^2~\epsilon_i(x)
\end{equation}
where we ignore the~$\hat{}$~on the quantum field. $A^2$ depends on the
static solution around which we expand. We label $A^2_V$ the operator for the
vacuum and $A^2_K$ the operator for the kink. We expand the quantum
fluctuation $\epsilon(t,x)$ in terms of the normal modes of the vacuum, which
we label $\epsilon_K(t,x)$, and the 1-kink, which we label
$\epsilon_V(t,x)$. In terms of the plane waves of the mesons with
eigenvalue $\omega_k$, 
\begin{equation}
\epsilon_V(t,x)= \sum_n\left[ \frac{e^{-i\omega_kt}}{\sqrt{2\omega_k}}
a(k)e^{ikx}~+~\frac{e^{i\omega_kt}}{\sqrt{2\omega_k}}a^{\dagger}(k)e^{-ikx}
\right]
\end{equation}
and, in terms of the normal modes of the 1-kink  with eigenvalue $\omega_n$,
\begin{equation}
\epsilon_K(t,x)=\sum_n\left[ \frac{e^{-i\omega_nt}}{\sqrt{2\omega_n}}a_n
\epsilon_n(x)~+~\frac{e^{i\omega_nt}}{\sqrt{2\omega_n}}
a_n^{\dagger}\epsilon_n^{\dagger}(x)\right]
\end{equation}
where the plane waves $\exp(ikx)$ and the normal modes $\epsilon_n$ are
orthonormal eigenfunctions. The next step involves writing the annihilation
and creation operators of the eigenmodes in terms of the creation and
annihilation operators of the planes waves. By definition,
\begin{eqnarray}
\epsilon_K(t,x)&=&\epsilon_V(t,x) \\
\dot{\epsilon}_K(t,x)&=&\dot{\epsilon}_V(t,x).
\end{eqnarray}
We then integrate over both equations with $x$, use the fact that the
eigenmodes are a complete orthonormal set (\ref{qrelations}) and solve the
set of two equations for $a_n$ and $a^{\dagger}_n$. We obtain
\begin{eqnarray}
a^{\dagger}_n=\frac{1}{2}\sum_k\left[ a^{\dagger}(k)\tilde{\epsilon}_n(-k)
\left(\sqrt{\frac{\omega_n}{\omega_k}}+\sqrt{\frac{\omega_k}{\omega_n}}\right)
+a^{\dagger}(k)\tilde{\epsilon}_n(k)
\left(\sqrt{\frac{\omega_n}{\omega_k}}-\sqrt{\frac{\omega_k}{\omega_n}}\right)
\right]\\ a_n=\frac{1}{2}\sum_k\left[ a^{\dagger}(k)\tilde{\epsilon}_n(-k)
\left(\sqrt{\frac{\omega_n}{\omega_k}}-\sqrt{\frac{\omega_k}{\omega_n}}\right)
+a^{\dagger}(k)\tilde{\epsilon}_n(k)
\left(\sqrt{\frac{\omega_n}{\omega_k}}+\sqrt{\frac{\omega_k}{\omega_n}}\right)
\right]
\end{eqnarray}
where $\tilde{\epsilon}_n(k)$ is the exponential Fourier transform, $\int
dx\exp(ikx)\epsilon_n(x)$, of $\epsilon_n(x)$. We calculate the hamiltonian
in terms of soliton normal modes annihilation and creation operators in the
last section (\ref{ham:norm}) and expand the term in terms of the
annihilation and creation operators of the vacuum. We get
\begin{eqnarray}
\omega_na^{\dagger}_na_n&=&\mbox{~`terms with $a^{\dagger}a$, $aa$ and
$a^{\dagger}a^{\dagger}$'~} \nonumber \\ &+&\frac{1}{4}\omega_n\sum_{k,l}
a(k)\tilde{\epsilon}_n(k)
\left(\sqrt{\frac{\omega_n}{\omega_k}}-\sqrt{\frac{\omega_k}{\omega_n}}\right)
a^{\dagger}(l)\tilde{\epsilon}_n(-l)
\left(\sqrt{\frac{\omega_n}{\omega_l}}-\sqrt{\frac{\omega_l}{\omega_n}}\right)
\nonumber\\ &=&\mbox{~`terms with $a^{\dagger}a$, $aa$ and
$a^{\dagger}a^{\dagger}$'~} \nonumber \\
&+&\frac{1}{4}\sum_{k}\tilde{\epsilon}_n(k)\tilde{\epsilon}_n(-k)
\left[\frac{\omega_n}{\omega_k}+\frac{\omega_k}{\omega_n}-2\right]\omega_n
\end{eqnarray}
where we have used the commutation relation between operators
$[a(k),a^{\dagger}(l)]=\delta_{k,l}$ and merged the resulting $a^{\dagger}a$
term into the `collective' term. Finally, we can express the un-ordered term
as the normal-ordered term i.e.\ all the terms with $a^{\dagger}a$, $aa$ and
$a^{\dagger}a^{\dagger}$ and an extra term:
\begin{equation}
\omega_na^{\dagger}_na_n~=~:\omega_na^{\dagger}_na_n:~+~\frac{1}{4}
\sum_{k}\tilde{\epsilon}_n(k)\tilde{\epsilon}_n(-k)
\frac{(\omega_n-\omega_k)^2}{\omega_k}
\end{equation}
which leads us to the final answer
\begin{equation}
:H:~=~:\omega_na^{\dagger}_na_n:~=~\omega_na^{\dagger}_na_n+\delta m.
\end{equation}
The mass correction $\delta m$ can be expressed as a trace over any complete
set of orthonormal states,
\begin{equation}\label{trace}
\delta m~=~-\frac{1}{4}Tr\left[\frac{(A_K-A_V)^2}{A_V}\right]
\end{equation}
where $A^2_V$ is the operator of the vacuum and $A^2_K$ is the
operator of the kink perturbations. This trace formula is finite. Further 
further discussions see \cite{bt} and \cite{barnes}.

\section{Trace formula: Theoretical Result}

We use the trace formula to calculate the contribution of the lowest discrete
modes to the mass correction. Cahill et al.\ only quote the results in their
paper \cite{trace:cahill}.\footnote{see also \cite{holz:trace} for explicit
calculations}  We can re-write the trace formula (\ref{trace}) in the
following way
\begin{equation}\label{general}
\delta m =-\frac{1}{4}\sum_n\int_{-\infty}^{\infty}dk~|\eta_{K,n}(x)
\eta_{V,k}(x)|^2~\left[\omega_n^2\omega_k^{-1}-2\omega_n+\omega_k\right]
\end{equation}
which reduces for the special case of the zero mode mass correction to
\begin{equation}\label{zero}
\delta m_0 =-\frac{1}{4}\int_{-\infty}^{\infty}dk~|\eta_{K,0}(x)
\eta_{V,k}(x)|^2~\omega_k
\end{equation}
where $\eta_{K,n}$ are the eigenmodes of the kink and $\eta_{V,k}$ the
eigenmodes of the vacuum. Finding the appropriate
Fourier transform is the main technical difficulty in solving these kinds of
integrals. We have used the Maple library {\it inttrans} \/to find Fourier
exponential, cos and sin transforms and the book on integral tables by
Erdelyi et al. \cite{inttable}. 

\subsection{$\phi^4$ kink model}

We have seen that the $\phi^4$-kink has two discrete modes (\ref{phi:modes}).
The zero mode
\begin{equation}
\eta_{K,0}(x)=\sqrt{\frac{3m}{4\sqrt{2}} }\cosh^{-2}
\left(\frac{mx}{\sqrt{2}}\right)
\end{equation}
with $\omega_{K,0}^2=0$ and the second discrete mode 
\begin{equation}
\eta_{K,1}(x)=\sqrt{\frac{3m}{2\sqrt{2}}}
\sinh\left(\frac{mx}{\sqrt{2}}\right)\cosh^{-2}\left(\frac{mx}{\sqrt{2}}\right)
\end{equation}
with $\omega_{K,1}^2=\frac{3}{2}m^2$ are here given in their normalised form
i.e.\ integration of the mode squared over $x$ gives one. The normalised
eigenmodes of the vacuum fluctuations are 
\begin{equation}
\eta_{V,k}(x)=\frac{1}{\sqrt{2\pi}}e^{ikx}
\end{equation}
with eigenvalues $\omega_{V,k}=k^2+2m^2$.  Using (\ref{zero}), we obtain the
following integral
\begin{equation}
\delta m~=~(\dots)\int_{-\infty}^{\infty}dk~\sqrt{k^2+2m^2} \left[\int dx
\frac{e^{ikx}}{\cosh^2\left(\frac{mx}{\sqrt{2}}\right)}\right] \left[\int dy
\frac{e^{-iky}}{\cosh^2\left(\frac{my}{\sqrt{2}}\right)}\right]
\end{equation}
which we simplify to 
\begin{eqnarray}
\delta m_0&=&(\dots)\int_{-\infty}^{\infty}dk~\sqrt{k^2+2m^2}
\left[\int_0^{\infty} dx\frac{\cos(kx)}{\cosh^2\left(\frac{mx}
{\sqrt{2}}\right)}\right]^2=(\dots)\int_{-\infty}^{\infty}
dk~\frac{\sqrt{k^2+2m^2}~k^2}{\sinh^2\left(\frac{k\pi}{\sqrt{2}m}\right)}
\nonumber\\ &=&-\frac{3m}{2\sqrt{2}\pi^3}\underbrace{\int_{-\infty}^{\infty}
dq~\frac{q^2\sqrt{q^2+\pi^2}}{\sinh^2q}}_{I_0}
\end{eqnarray}
where we have used Euler's formula, the Fourier cos transform of $\cosh^{-2}$
and changed to the variable $q\equiv\frac{k\pi}{\sqrt{2}m}$.  Using
(\ref{general}), we obtain the correction to the mass from the second
discrete mode
\begin{eqnarray}
\delta m_1&=&(\dots)\int_{-\infty}^{\infty}dk
\left(\frac{3m^2}{2}(k^2+2m^2)^{-\frac{1}{2}}-3m^2+\sqrt{k^2+2m^2}\right)
\left[\int_0^{\infty}dx\sin(kx)\frac{\sinh\left(\frac{mx}{\sqrt{2}}\right)}
{\cosh^2\left(\frac{mx}{\sqrt{2}}\right)}\right]^2\nonumber\\
&=&(\dots)\int_{-\infty}^{\infty}dk~(\dots)
~\frac{k^2\cosh^2\left(\frac{mx}{\sqrt{2}}\right)}
{\left(1+\cosh^2\left(\frac{mx}{\sqrt{2}}\right)\right)}\nonumber\\
&=&-\left(\frac{1}{4}\right)\frac{3m}{2\sqrt{2}\pi^3}
\underbrace{\int_{-\infty}^{\infty}dq~\left(\frac{(\sqrt{q^2+4\pi}
-\sqrt{3\pi^2})^2}{\sqrt{q^2+4\pi^2}}\right) \frac{q^2}{\left[1+\cosh
q\right]^2}}_{I_1}
\end{eqnarray}
where we have used Euler's formula, the Fourier sin transform of
$\sinh\cosh^{-2}$ and changed to the variable $q\equiv\frac{k\pi}{\sqrt{2}m}$.

We have evaluated $I_0$ and $I_1$ numerically with Maple and obtain 
\begin{eqnarray}
I_0=11.247\nonumber\\ \frac{1}{4}I_1=0.827\nonumber
\end{eqnarray}
and the contributions to the mass correction are 
\begin{equation}\label{phi4_lowest}
\delta m_0+\delta m_1~=~(-0.384-0.0283)m~=~-0.413m
\end{equation}
compared to the full quantum correction (\ref{full})
\begin{equation}
\delta m~=~-0.471m
\end{equation}
where we have set $\hbar$ to one. Finally, we find that the zero mode 
contributes $81.5\%$ and the second discrete mode $6\%$, thus in total 
$87.5\%$, to the total mass correction. 

We are also interested to what value of $k$ the normal modes of the vacuum
fluctuations have to go to give a reliable answer to the mass
correction. Intuitively, the zero and second discrete mode are localised and
its norm with the long wavelength vacuum modes should become very small.  We
have put a cut-off $\Lambda$ on our integral $I_0$ and $\frac{1}{4}I_1$ and
evaluated the integrals as a function of the cut-off. We have done this
numerically with Maple. Figure~\ref{phi:k:dependence} shows that we only need
to go up to a cut-off of around $q=5$ for the zero mode and of around $q=15$
for the second discrete mode. (k$\equiv\frac{\sqrt{2}m}{\pi}q$)
This is good news, for we can ignore long
wavelength vacuum modes. 
\begin{figure}
\begin{center}
\epsfig{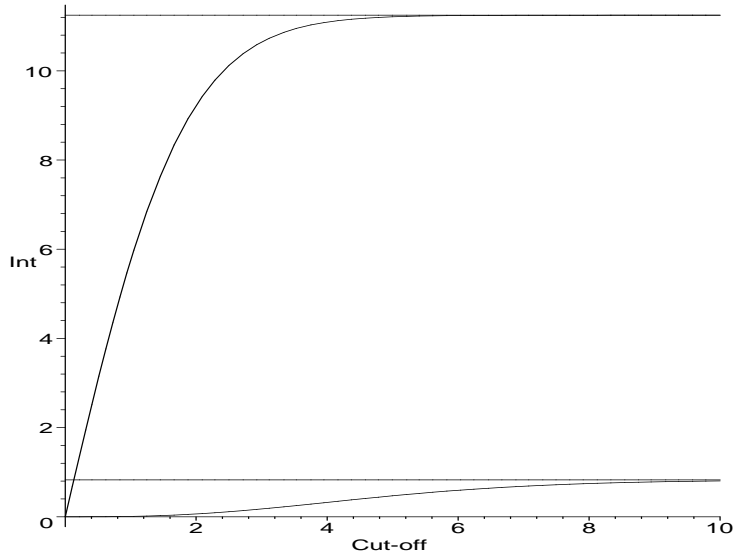}
\caption{Value of $I_0$ (upper curve) and $I_1$ (lower curve) as a 
function of their cut-off $\Lambda$ for q}
\label{phi:k:dependence}
\end{center}
\end{figure}

\subsection{Sine-Gordon model}

We do the same for the Sine-Gordon model. There is only one discrete mode,
the zero mode of translation: see section~\ref{jack} and \cite{jackiw}. The
normalised zero mode has the form
\begin{equation}
\eta_{K,0}(x)=\sqrt{\frac{m}{2}}\cosh^{-1}(mx)
\end{equation}
with eigenvalue $\omega_{K,0}=0$. The eigenmodes of the vacuum fluctuations
are the same as before, but the eigenvalues change to
$\omega_{V,k}=k^2+m^2$. The mass correction takes the form
\begin{eqnarray}
\delta m_0&=&(\dots)\int dk \sqrt{k^2+m^2}~\left[\int dx \cos(kx)
\cosh^{-1}(mx)\right]^2\nonumber\\ &=&(\dots)\int dk
\sqrt{k^2+m^2}~\cosh^{-2}\left(\frac{\pi k}{2m}\right)\nonumber\\
&=&-\frac{m}{4\pi}\int dq
\frac{\sqrt{q^2+\left(\frac{\pi}{2}\right)^2}}{\cosh^2q}
\end{eqnarray}
where we have used Euler's formula, the Fourier cos transform of $\cosh^{-1}$
and changed to the variable $q\equiv\frac{k\pi}{2m}$.

We have evaluated the integral numerically with Maple and obtain
\begin{equation}\label{sine_lowest}
\delta m_0~=~-3.572\frac{m}{4\pi}~=~-0.893~\frac{m}{\pi}~=~-0.284~m
\end{equation}
compared to the full quantum correction (\cite{dashen}, \cite{raj}) 
\begin{equation}
\delta m~=~-\frac{m}{\pi}~=~-0.318~m
\end{equation}
where we have set $\hbar$  to one. Finally, we find that the zero mode contributes
$89.3\%$ to the total mass correction. 

Again, we are interested to what value of $k$ the normal modes of the vacuum
fluctuations have to go to give a reliable answer to the mass correction. We
have put a cut-off $\Lambda$ on the integral and evaluated the integral as a
function of the cut-off. We have done this numerically with
Maple. Figure~\ref{sine:k:dependence} shows that we only need to go up to a
cut-off of around $q=5$ for the zero mode. 
\begin{figure}
\begin{center}
\epsfig{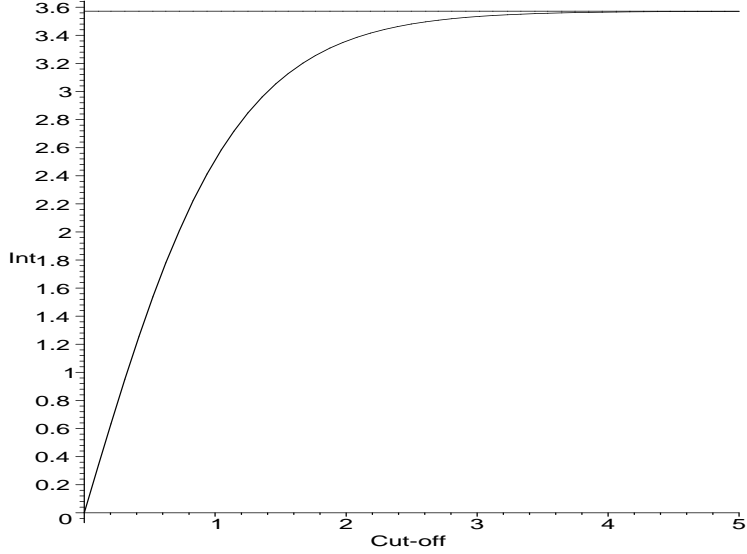}
\caption{The value of the integral versus its cut-off $\Lambda$ for $q$}
\label{sine:k:dependence}
\end{center}
\end{figure}
We do not need to include long wavelengths vacuum modes. 

\section{Trace formula: Numerical Result}

In the last section, we have applied the trace formula to the $\phi^4$ kink
and the Sine-Gordon model. The results are clear-cut. The contribution from
the discrete modes to the quantum mass correction is dominant (more than
$80\%$). Further, we do not need to probe our discrete modes for long
wavelength of the vacuum mode. This is good news for numerical methods and we
can limit ourselves to the lowest normal modes of fluctuations in both the
vacuum and the kink sector. Moreover, in (1+1) dimension, we are not
restricted by memory or computational needs and can include all the vacuum
and kink modes. 

\subsection{Preparation}

We calculate the mass correction for the $\phi^4$ kink and Sine-Gordon
model. We have set the mass $m=1$ and coupling $\lambda=1$ for simplicity.  
The energy functional has the form
\begin{equation}
E=\int dx\left[\frac{1}{2}\dot{\phi}^2+\frac{1}{2}{\phi'}^2+
\frac{1}{4}(\phi^2-1)^2\right]
\end{equation}
for the $\phi^4$ kink and 
\begin{equation}
E=\int dx\left[\frac{1}{2}\dot{\phi}^2+\frac{1}{2}{\phi'}^2+
(1-\cos\phi)\right]
\end{equation}
for the Sine-Gordon model; where $\phi'=\frac{d\phi}{dx}$ and 
$\dot{\phi}=\frac{d\phi}{dt}$. Using appropriate boundary conditions, 
we find the
minimal-energy configuration, which we call $\phi_{st}$, for both models in
the topological charge sector one. We use the Gauss-Seidel overrelaxation
method. Our box size is $L=40$ from -20 to 20 and we use 1600 points. Thus,
the lattice spacing is $dx=0.025$. 

The corresponding eigenvalue vacuum and kink operators in terms of the 
static solution 
(see \ref{qhamilton}) are
\begin{eqnarray}
A^2_V&=&-\frac{d^2}{dx^2}+2 \nonumber\\
A^2_K&=&-\frac{d^2}{dx^2}+1-3\phi_{st}^2
\end{eqnarray}
for the $\phi^4$ kink and 
\begin{eqnarray}
A^2_V&=&-\frac{d^2}{dx^2}+1 \nonumber\\
A^2_K&=&-\frac{d^2}{dx^2}+\cos\phi_{st}
\end{eqnarray}
for the Sine-Gordon model. In {\it Numerical Techniques}, we describe three
different methods that solve the discrete eigenvalue problem. The trivial
matrix diagonalisation is the more accurate and simplest one. However, we
have to admit that the computational time grows as the cube of the number of points 
and the technique cannot be used in two or more dimensions. We substitute the
value of the numerically minimised field\footnote{we used the relaxation
  method described in {\it Numerical Methods}} into the discretised eigenvalue equation and
diagonalise the resulting matrix with periodic boundary conditions. 

Figure \ref{Phi4_modes0-3} shows the first four normal modes of the $\phi^4$
kink. 
\begin{figure}
\begin{center}
\includegraphics[angle=-90,scale=1,width=0.7\textwidth]{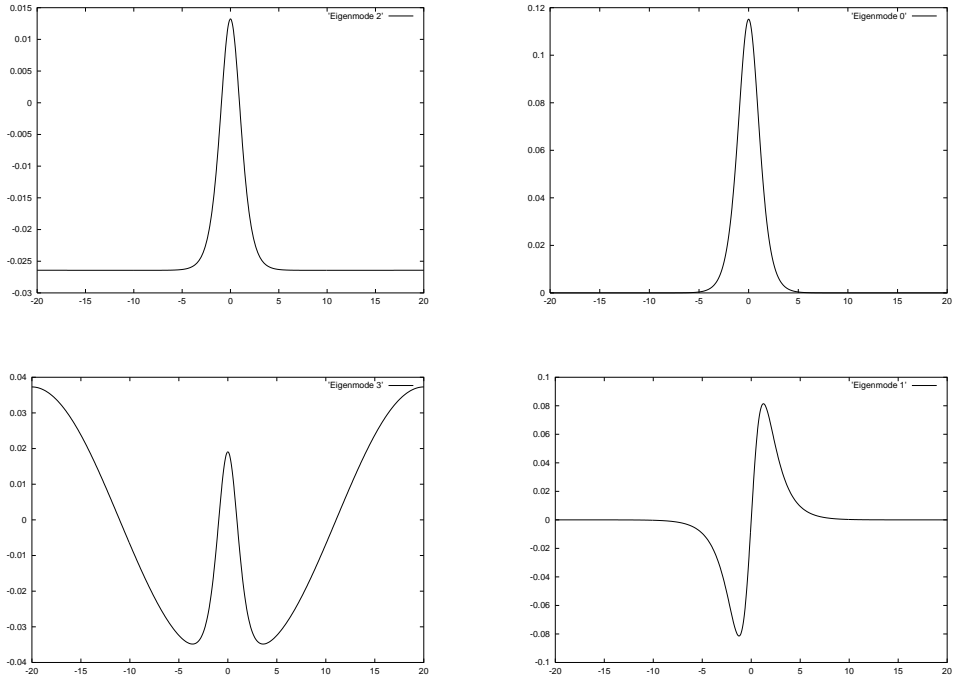}
\caption{The $\phi^4$ Kink eigenmodes from zero to three}
\label{Phi4_modes0-3}
\end{center}
\end{figure}
\begin{figure}
\begin{center}
\includegraphics[angle=-90,scale=1,width=0.7\textwidth]{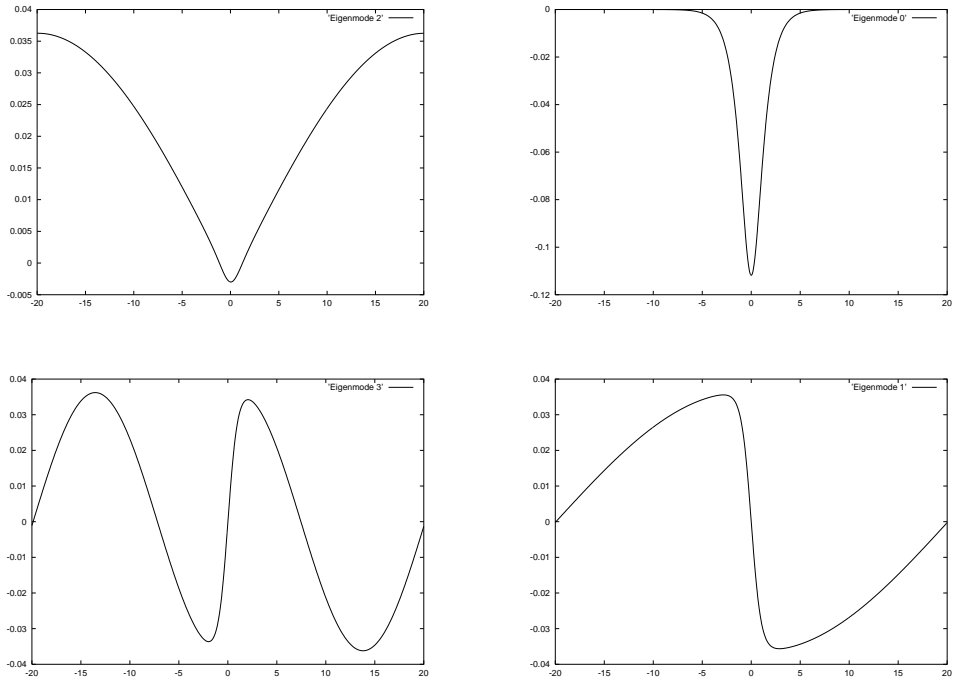}
\caption{The Sine-Gordon eigenmodes from zero to three}
\label{Sine_modes0-3}
\end{center}
\end{figure}
The numerical eigenvalues are $-5.78~10^{-8}$, 1.49998, 2 and 2.03072
compared to the exact eigenvalues 0 (zero mode), 1.5 (first mode) and 2
(start of `continuous' modes).  Figure \ref{Sine_modes0-3} shows the first
four normal modes of the Sine-Gordon soliton. The numerical eigenvalues are
$5.12~10^{-8}$, 1.00682, 1.00682 and 1.06128 compared to the exact
eigenvalues 0 (zero mode) and 1 (start of `continuous' modes).

\subsection{Results}

We have all the information needed to compute the mass correction. The trace
formula has the form 
\begin{equation}
\delta m =-\frac{1}{4}\sum_{n,k}
\left(\sum_i\eta_{K,n}(x_i)\eta_{V,k}(x_i)\right)^2
~\left[\omega_n^2\omega_k^{-1}-2\omega_n+\omega_k\right]
\end{equation}
which is similar to (\ref{general}). $\eta_{K,n}(x_i)$ refers to the nth
eigenmode of the kink, $\eta_{V,k}(x_i)$ refers to the kth eigenmode of the
vacuum and $x_i$ to the position of the ith lattice point.

We start with the $\phi^4$ kink. We sum up the contributions to the mass
correction, mode by mode. Figure  \ref{Phi4_total_correction} shows that the
mass correction approaches an asymptotic limit. 
\begin{figure}
\begin{center}
\input{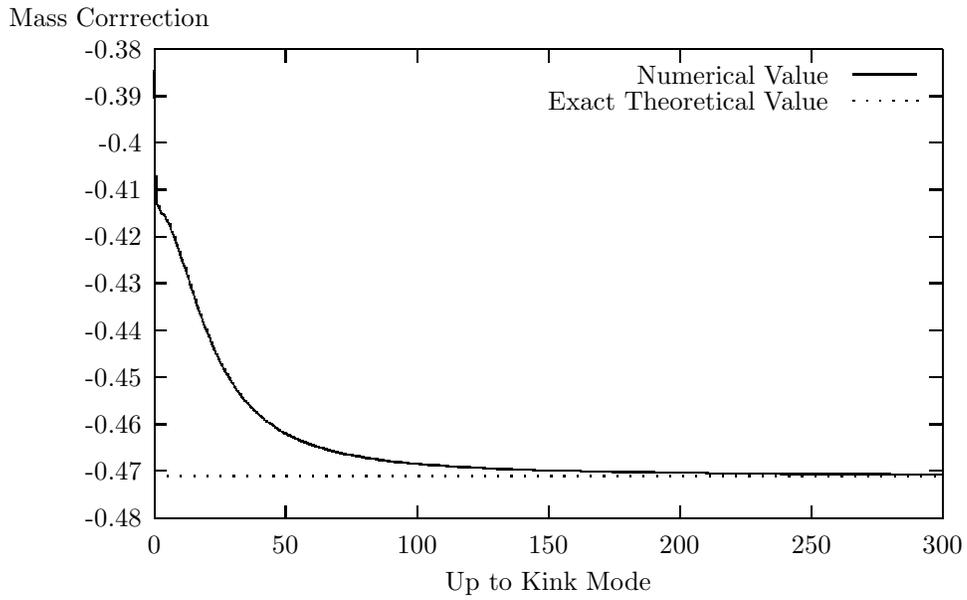}
\end{center}
\caption{The total contribution to the mass correction mode by mode}
\label{Phi4_total_correction}
\end{figure}
The first few mode contributions are the most important ones. We
only include mode contributions up to mode 200. Figure
\ref{Phi4_modes_contribution} shows that the norm is mostly zero and peaks for
the norm between the nth kink mode and the nth vacuum mode. For the lowest kink
mode contribution, only the lowest vacuum modes are important. For the
highest kink mode contributions, only the highest vacuum modes are important.
\begin{figure}
\begin{center}
\setlength{\unitlength}{0.240900pt}
\ifx\plotpoint\undefined\newsavebox{\plotpoint}\fi
\sbox{\plotpoint}{\rule[-0.200pt]{0.400pt}{0.400pt}}%
\begin{picture}(1500,900)(0,0)
\font\gnuplot=cmr10 at 10pt
\gnuplot
\sbox{\plotpoint}{\rule[-0.200pt]{0.400pt}{0.400pt}}%
\put(161.0,123.0){\rule[-0.200pt]{4.818pt}{0.400pt}}
\put(141,123){\makebox(0,0)[r]{0}}
\put(1419.0,123.0){\rule[-0.200pt]{4.818pt}{0.400pt}}
\put(161.0,197.0){\rule[-0.200pt]{4.818pt}{0.400pt}}
\put(141,197){\makebox(0,0)[r]{0.1}}
\put(1419.0,197.0){\rule[-0.200pt]{4.818pt}{0.400pt}}
\put(161.0,270.0){\rule[-0.200pt]{4.818pt}{0.400pt}}
\put(141,270){\makebox(0,0)[r]{0.2}}
\put(1419.0,270.0){\rule[-0.200pt]{4.818pt}{0.400pt}}
\put(161.0,344.0){\rule[-0.200pt]{4.818pt}{0.400pt}}
\put(141,344){\makebox(0,0)[r]{0.3}}
\put(1419.0,344.0){\rule[-0.200pt]{4.818pt}{0.400pt}}
\put(161.0,418.0){\rule[-0.200pt]{4.818pt}{0.400pt}}
\put(141,418){\makebox(0,0)[r]{0.4}}
\put(1419.0,418.0){\rule[-0.200pt]{4.818pt}{0.400pt}}
\put(161.0,492.0){\rule[-0.200pt]{4.818pt}{0.400pt}}
\put(141,492){\makebox(0,0)[r]{0.5}}
\put(1419.0,492.0){\rule[-0.200pt]{4.818pt}{0.400pt}}
\put(161.0,565.0){\rule[-0.200pt]{4.818pt}{0.400pt}}
\put(141,565){\makebox(0,0)[r]{0.6}}
\put(1419.0,565.0){\rule[-0.200pt]{4.818pt}{0.400pt}}
\put(161.0,639.0){\rule[-0.200pt]{4.818pt}{0.400pt}}
\put(141,639){\makebox(0,0)[r]{0.7}}
\put(1419.0,639.0){\rule[-0.200pt]{4.818pt}{0.400pt}}
\put(161.0,713.0){\rule[-0.200pt]{4.818pt}{0.400pt}}
\put(141,713){\makebox(0,0)[r]{0.8}}
\put(1419.0,713.0){\rule[-0.200pt]{4.818pt}{0.400pt}}
\put(161.0,786.0){\rule[-0.200pt]{4.818pt}{0.400pt}}
\put(141,786){\makebox(0,0)[r]{0.9}}
\put(1419.0,786.0){\rule[-0.200pt]{4.818pt}{0.400pt}}
\put(161.0,860.0){\rule[-0.200pt]{4.818pt}{0.400pt}}
\put(141,860){\makebox(0,0)[r]{1}}
\put(1419.0,860.0){\rule[-0.200pt]{4.818pt}{0.400pt}}
\put(344.0,123.0){\rule[-0.200pt]{0.400pt}{4.818pt}}
\put(310,600){\makebox(0,0){Mode 50}} 
\put(344,82){\makebox(0,0){200}}
\put(344.0,840.0){\rule[-0.200pt]{0.400pt}{4.818pt}}
\put(587.0,123.0){\rule[-0.200pt]{0.400pt}{4.818pt}}
\put(500,400){\makebox(0,0){Mode 250}} 
\put(587,82){\makebox(0,0){400}}
\put(587.0,840.0){\rule[-0.200pt]{0.400pt}{4.818pt}}
\put(830.0,123.0){\rule[-0.200pt]{0.400pt}{4.818pt}}
\put(720,750){\makebox(0,0){Mode 500}} 
\put(830,82){\makebox(0,0){600}}
\put(830.0,840.0){\rule[-0.200pt]{0.400pt}{4.818pt}}
\put(1074.0,123.0){\rule[-0.200pt]{0.400pt}{4.818pt}}
\put(1074,82){\makebox(0,0){800}}
\put(1074.0,840.0){\rule[-0.200pt]{0.400pt}{4.818pt}}
\put(1317.0,123.0){\rule[-0.200pt]{0.400pt}{4.818pt}}
\put(1317,82){\makebox(0,0){1000}}
\put(1300,700){\makebox(0,0){Mode 1000}} 
\put(1317.0,840.0){\rule[-0.200pt]{0.400pt}{4.818pt}}
\put(161.0,123.0){\rule[-0.200pt]{307.870pt}{0.400pt}}
\put(1439.0,123.0){\rule[-0.200pt]{0.400pt}{177.543pt}}
\put(161.0,860.0){\rule[-0.200pt]{307.870pt}{0.400pt}}
\put(120,900){\makebox(0,0){Norm with Kink $\dots$}}
\put(800,21){\makebox(0,0){and Vacuum Mode}}
\put(161.0,123.0){\rule[-0.200pt]{0.400pt}{177.543pt}}
\put(161,123){\usebox{\plotpoint}}
\put(161.0,123.0){\rule[-0.200pt]{307.870pt}{0.400pt}}
\put(161,123){\usebox{\plotpoint}}
\put(216,122.67){\rule{0.241pt}{0.400pt}}
\multiput(216.00,122.17)(0.500,1.000){2}{\rule{0.120pt}{0.400pt}}
\put(217,122.67){\rule{0.241pt}{0.400pt}}
\multiput(217.00,123.17)(0.500,-1.000){2}{\rule{0.120pt}{0.400pt}}
\put(217.67,123){\rule{0.400pt}{1.445pt}}
\multiput(217.17,123.00)(1.000,3.000){2}{\rule{0.400pt}{0.723pt}}
\put(219,128.67){\rule{0.482pt}{0.400pt}}
\multiput(219.00,128.17)(1.000,1.000){2}{\rule{0.241pt}{0.400pt}}
\put(220.67,130){\rule{0.400pt}{170.798pt}}
\multiput(220.17,130.00)(1.000,354.500){2}{\rule{0.400pt}{85.399pt}}
\put(221.67,123){\rule{0.400pt}{172.484pt}}
\multiput(221.17,481.00)(1.000,-358.000){2}{\rule{0.400pt}{86.242pt}}
\put(222.67,123){\rule{0.400pt}{0.964pt}}
\multiput(222.17,123.00)(1.000,2.000){2}{\rule{0.400pt}{0.482pt}}
\put(224.17,123){\rule{0.400pt}{0.900pt}}
\multiput(223.17,125.13)(2.000,-2.132){2}{\rule{0.400pt}{0.450pt}}
\put(226,122.67){\rule{0.241pt}{0.400pt}}
\multiput(226.00,122.17)(0.500,1.000){2}{\rule{0.120pt}{0.400pt}}
\put(227,122.67){\rule{0.241pt}{0.400pt}}
\multiput(227.00,123.17)(0.500,-1.000){2}{\rule{0.120pt}{0.400pt}}
\put(161.0,123.0){\rule[-0.200pt]{13.249pt}{0.400pt}}
\put(228.0,123.0){\rule[-0.200pt]{291.730pt}{0.400pt}}
\put(161,123){\usebox{\plotpoint}}
\put(400,122.67){\rule{0.241pt}{0.400pt}}
\multiput(400.00,122.17)(0.500,1.000){2}{\rule{0.120pt}{0.400pt}}
\put(401,122.67){\rule{0.241pt}{0.400pt}}
\multiput(401.00,123.17)(0.500,-1.000){2}{\rule{0.120pt}{0.400pt}}
\put(401.67,123){\rule{0.400pt}{166.462pt}}
\multiput(401.17,123.00)(1.000,345.500){2}{\rule{0.400pt}{83.231pt}}
\put(402.67,166){\rule{0.400pt}{156.103pt}}
\multiput(402.17,490.00)(1.000,-324.000){2}{\rule{0.400pt}{78.052pt}}
\put(404.17,124){\rule{0.400pt}{8.500pt}}
\multiput(403.17,148.36)(2.000,-24.358){2}{\rule{0.400pt}{4.250pt}}
\put(406,122.67){\rule{0.241pt}{0.400pt}}
\multiput(406.00,123.17)(0.500,-1.000){2}{\rule{0.120pt}{0.400pt}}
\put(161.0,123.0){\rule[-0.200pt]{57.575pt}{0.400pt}}
\put(407.0,123.0){\rule[-0.200pt]{248.609pt}{0.400pt}}
\put(161,123){\usebox{\plotpoint}}
\put(705.67,123){\rule{0.400pt}{138.036pt}}
\multiput(705.17,123.00)(1.000,286.500){2}{\rule{0.400pt}{69.018pt}}
\put(707.17,286){\rule{0.400pt}{82.100pt}}
\multiput(706.17,525.60)(2.000,-239.597){2}{\rule{0.400pt}{41.050pt}}
\put(708.67,123){\rule{0.400pt}{39.267pt}}
\multiput(708.17,204.50)(1.000,-81.500){2}{\rule{0.400pt}{19.633pt}}
\put(161.0,123.0){\rule[-0.200pt]{131.290pt}{0.400pt}}
\put(710.0,123.0){\rule[-0.200pt]{175.616pt}{0.400pt}}
\put(161,123){\usebox{\plotpoint}}
\put(1314.67,123){\rule{0.400pt}{54.925pt}}
\multiput(1314.17,123.00)(1.000,114.000){2}{\rule{0.400pt}{27.463pt}}
\put(1315.67,351){\rule{0.400pt}{67.693pt}}
\multiput(1315.17,351.00)(1.000,140.500){2}{\rule{0.400pt}{33.846pt}}
\put(1317.17,123){\rule{0.400pt}{101.900pt}}
\multiput(1316.17,420.50)(2.000,-297.501){2}{\rule{0.400pt}{50.950pt}}
\put(161.0,123.0){\rule[-0.200pt]{277.999pt}{0.400pt}}
\put(1319.0,123.0){\rule[-0.200pt]{28.908pt}{0.400pt}}
\end{picture}
\end{center}
\caption{Norm between four kink modes and the vacuum modes}
\label{Phi4_modes_contribution}
\end{figure}
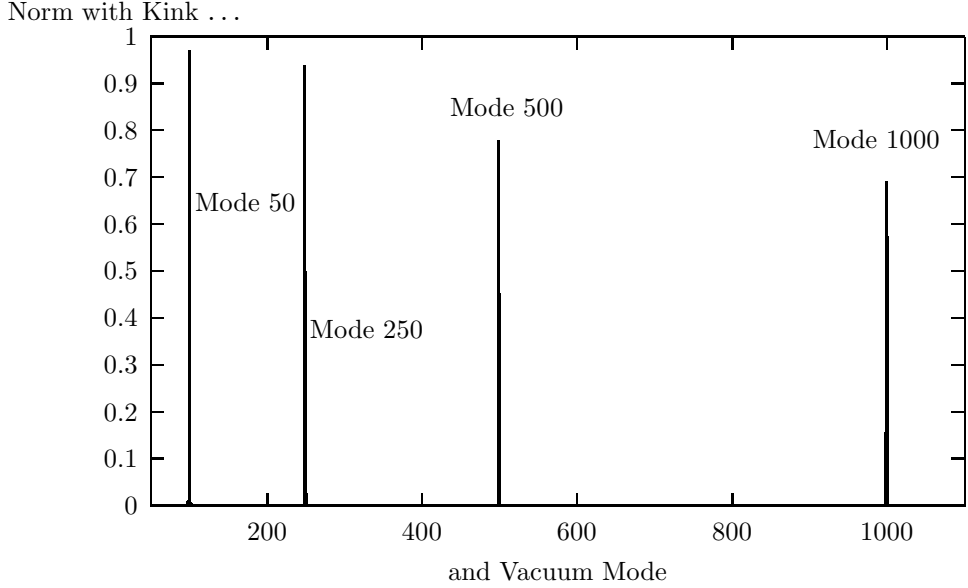
The discrete mode corrections are 
\begin{eqnarray}
\delta m_0&=&-0.384626 \nonumber \\
\delta m_1&=&-0.0282964
\end{eqnarray}
which are very close to their exact values (\ref{phi4_lowest}). The numerical
value of the mass correction is 
\begin{equation}
\delta m~=~-0.471097
\end{equation}
compared to the exact value of -0.471113 and is 99.997\% accurate. This is 
a very satisfactory result. 

We turn our attention to the Sine-Gordon kink. We sum up the contributions 
to the mass correction mode by mode. Figure 
\ref{Sine_total_correction} shows that the mass correction approaches an
asymptotic limit. 
\begin{figure}
\begin{center}
\input{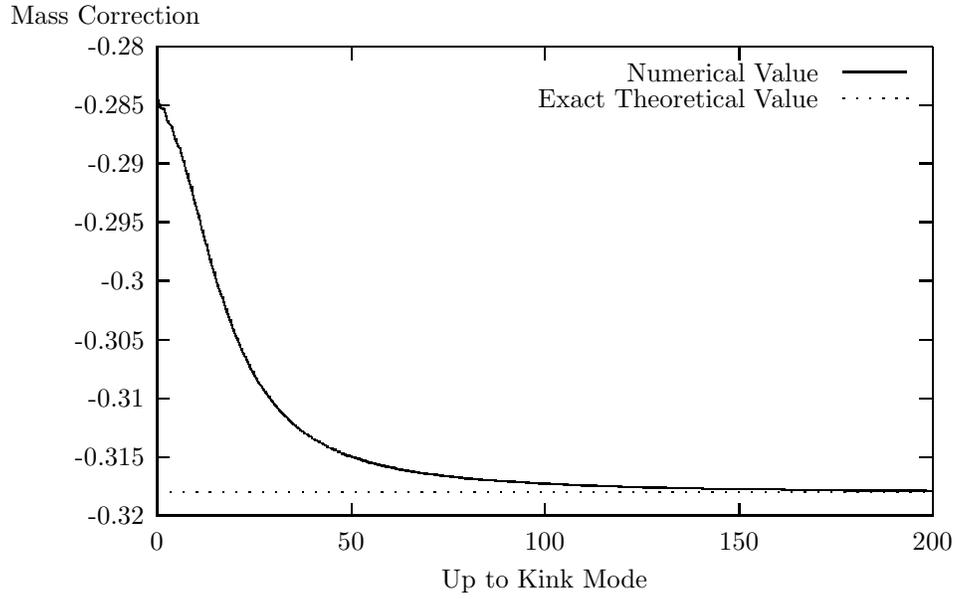}
\end{center}
\caption{The total contribution to the mass correction mode by mode}
\label{Sine_total_correction}
\end{figure}
The first few mode contributions are the most important ones. Figure 
\ref{Sine_mode1_contribution} shows the contribution of the first mode for
each vacuum mode. Note that some contributions are zero, because the first
mode is an odd function and some of the vacuum modes are even functions. 
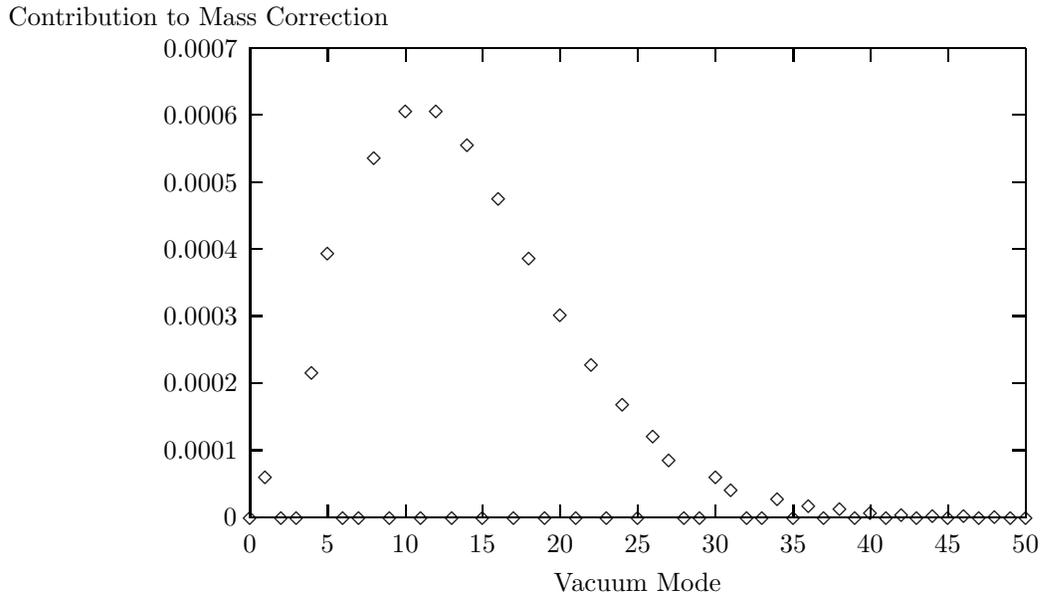
\begin{figure}
\begin{center}
\setlength{\unitlength}{0.240900pt}
\ifx\plotpoint\undefined\newsavebox{\plotpoint}\fi
\sbox{\plotpoint}{\rule[-0.200pt]{0.400pt}{0.400pt}}%
\begin{picture}(1500,900)(0,0)
\font\gnuplot=cmr10 at 10pt
\gnuplot
\sbox{\plotpoint}{\rule[-0.200pt]{0.400pt}{0.400pt}}%
\put(221.0,123.0){\rule[-0.200pt]{4.818pt}{0.400pt}}
\put(201,123){\makebox(0,0)[r]{0}}
\put(1419.0,123.0){\rule[-0.200pt]{4.818pt}{0.400pt}}
\put(221.0,228.0){\rule[-0.200pt]{4.818pt}{0.400pt}}
\put(201,228){\makebox(0,0)[r]{0.0001}}
\put(1419.0,228.0){\rule[-0.200pt]{4.818pt}{0.400pt}}
\put(221.0,334.0){\rule[-0.200pt]{4.818pt}{0.400pt}}
\put(201,334){\makebox(0,0)[r]{0.0002}}
\put(1419.0,334.0){\rule[-0.200pt]{4.818pt}{0.400pt}}
\put(221.0,439.0){\rule[-0.200pt]{4.818pt}{0.400pt}}
\put(201,439){\makebox(0,0)[r]{0.0003}}
\put(1419.0,439.0){\rule[-0.200pt]{4.818pt}{0.400pt}}
\put(221.0,544.0){\rule[-0.200pt]{4.818pt}{0.400pt}}
\put(201,544){\makebox(0,0)[r]{0.0004}}
\put(1419.0,544.0){\rule[-0.200pt]{4.818pt}{0.400pt}}
\put(221.0,649.0){\rule[-0.200pt]{4.818pt}{0.400pt}}
\put(201,649){\makebox(0,0)[r]{0.0005}}
\put(1419.0,649.0){\rule[-0.200pt]{4.818pt}{0.400pt}}
\put(221.0,755.0){\rule[-0.200pt]{4.818pt}{0.400pt}}
\put(201,755){\makebox(0,0)[r]{0.0006}}
\put(1419.0,755.0){\rule[-0.200pt]{4.818pt}{0.400pt}}
\put(221.0,860.0){\rule[-0.200pt]{4.818pt}{0.400pt}}
\put(201,860){\makebox(0,0)[r]{0.0007}}
\put(1419.0,860.0){\rule[-0.200pt]{4.818pt}{0.400pt}}
\put(221.0,123.0){\rule[-0.200pt]{0.400pt}{4.818pt}}
\put(221,82){\makebox(0,0){0}}
\put(221.0,840.0){\rule[-0.200pt]{0.400pt}{4.818pt}}
\put(343.0,123.0){\rule[-0.200pt]{0.400pt}{4.818pt}}
\put(343,82){\makebox(0,0){5}}
\put(343.0,840.0){\rule[-0.200pt]{0.400pt}{4.818pt}}
\put(465.0,123.0){\rule[-0.200pt]{0.400pt}{4.818pt}}
\put(465,82){\makebox(0,0){10}}
\put(465.0,840.0){\rule[-0.200pt]{0.400pt}{4.818pt}}
\put(586.0,123.0){\rule[-0.200pt]{0.400pt}{4.818pt}}
\put(586,82){\makebox(0,0){15}}
\put(586.0,840.0){\rule[-0.200pt]{0.400pt}{4.818pt}}
\put(708.0,123.0){\rule[-0.200pt]{0.400pt}{4.818pt}}
\put(708,82){\makebox(0,0){20}}
\put(708.0,840.0){\rule[-0.200pt]{0.400pt}{4.818pt}}
\put(830.0,123.0){\rule[-0.200pt]{0.400pt}{4.818pt}}
\put(830,82){\makebox(0,0){25}}
\put(830.0,840.0){\rule[-0.200pt]{0.400pt}{4.818pt}}
\put(952.0,123.0){\rule[-0.200pt]{0.400pt}{4.818pt}}
\put(952,82){\makebox(0,0){30}}
\put(952.0,840.0){\rule[-0.200pt]{0.400pt}{4.818pt}}
\put(1074.0,123.0){\rule[-0.200pt]{0.400pt}{4.818pt}}
\put(1074,82){\makebox(0,0){35}}
\put(1074.0,840.0){\rule[-0.200pt]{0.400pt}{4.818pt}}
\put(1195.0,123.0){\rule[-0.200pt]{0.400pt}{4.818pt}}
\put(1195,82){\makebox(0,0){40}}
\put(1195.0,840.0){\rule[-0.200pt]{0.400pt}{4.818pt}}
\put(1317.0,123.0){\rule[-0.200pt]{0.400pt}{4.818pt}}
\put(1317,82){\makebox(0,0){45}}
\put(1317.0,840.0){\rule[-0.200pt]{0.400pt}{4.818pt}}
\put(1439.0,123.0){\rule[-0.200pt]{0.400pt}{4.818pt}}
\put(1439,82){\makebox(0,0){50}}
\put(1439.0,840.0){\rule[-0.200pt]{0.400pt}{4.818pt}}
\put(221.0,123.0){\rule[-0.200pt]{293.416pt}{0.400pt}}
\put(1439.0,123.0){\rule[-0.200pt]{0.400pt}{177.543pt}}
\put(221.0,860.0){\rule[-0.200pt]{293.416pt}{0.400pt}}
\put(140,910){\makebox(0,0){Contribution to Mass Correction}}
\put(830,21){\makebox(0,0){Vacuum Mode}}
\put(221.0,123.0){\rule[-0.200pt]{0.400pt}{177.543pt}}
\put(221,123){\raisebox{-.8pt}{\makebox(0,0){$\diamond$}}}
\put(245,188){\raisebox{-.8pt}{\makebox(0,0){$\diamond$}}}
\put(270,123){\raisebox{-.8pt}{\makebox(0,0){$\diamond$}}}
\put(294,123){\raisebox{-.8pt}{\makebox(0,0){$\diamond$}}}
\put(318,351){\raisebox{-.8pt}{\makebox(0,0){$\diamond$}}}
\put(343,540){\raisebox{-.8pt}{\makebox(0,0){$\diamond$}}}
\put(367,123){\raisebox{-.8pt}{\makebox(0,0){$\diamond$}}}
\put(392,123){\raisebox{-.8pt}{\makebox(0,0){$\diamond$}}}
\put(416,689){\raisebox{-.8pt}{\makebox(0,0){$\diamond$}}}
\put(440,123){\raisebox{-.8pt}{\makebox(0,0){$\diamond$}}}
\put(465,763){\raisebox{-.8pt}{\makebox(0,0){$\diamond$}}}
\put(489,123){\raisebox{-.8pt}{\makebox(0,0){$\diamond$}}}
\put(513,763){\raisebox{-.8pt}{\makebox(0,0){$\diamond$}}}
\put(538,123){\raisebox{-.8pt}{\makebox(0,0){$\diamond$}}}
\put(562,709){\raisebox{-.8pt}{\makebox(0,0){$\diamond$}}}
\put(586,123){\raisebox{-.8pt}{\makebox(0,0){$\diamond$}}}
\put(611,625){\raisebox{-.8pt}{\makebox(0,0){$\diamond$}}}
\put(635,123){\raisebox{-.8pt}{\makebox(0,0){$\diamond$}}}
\put(659,532){\raisebox{-.8pt}{\makebox(0,0){$\diamond$}}}
\put(684,123){\raisebox{-.8pt}{\makebox(0,0){$\diamond$}}}
\put(708,442){\raisebox{-.8pt}{\makebox(0,0){$\diamond$}}}
\put(733,123){\raisebox{-.8pt}{\makebox(0,0){$\diamond$}}}
\put(757,365){\raisebox{-.8pt}{\makebox(0,0){$\diamond$}}}
\put(781,123){\raisebox{-.8pt}{\makebox(0,0){$\diamond$}}}
\put(806,301){\raisebox{-.8pt}{\makebox(0,0){$\diamond$}}}
\put(830,123){\raisebox{-.8pt}{\makebox(0,0){$\diamond$}}}
\put(854,252){\raisebox{-.8pt}{\makebox(0,0){$\diamond$}}}
\put(879,214){\raisebox{-.8pt}{\makebox(0,0){$\diamond$}}}
\put(903,123){\raisebox{-.8pt}{\makebox(0,0){$\diamond$}}}
\put(927,123){\raisebox{-.8pt}{\makebox(0,0){$\diamond$}}}
\put(952,187){\raisebox{-.8pt}{\makebox(0,0){$\diamond$}}}
\put(976,167){\raisebox{-.8pt}{\makebox(0,0){$\diamond$}}}
\put(1001,123){\raisebox{-.8pt}{\makebox(0,0){$\diamond$}}}
\put(1025,123){\raisebox{-.8pt}{\makebox(0,0){$\diamond$}}}
\put(1049,153){\raisebox{-.8pt}{\makebox(0,0){$\diamond$}}}
\put(1074,123){\raisebox{-.8pt}{\makebox(0,0){$\diamond$}}}
\put(1098,143){\raisebox{-.8pt}{\makebox(0,0){$\diamond$}}}
\put(1122,123){\raisebox{-.8pt}{\makebox(0,0){$\diamond$}}}
\put(1147,137){\raisebox{-.8pt}{\makebox(0,0){$\diamond$}}}
\put(1171,123){\raisebox{-.8pt}{\makebox(0,0){$\diamond$}}}
\put(1195,132){\raisebox{-.8pt}{\makebox(0,0){$\diamond$}}}
\put(1220,123){\raisebox{-.8pt}{\makebox(0,0){$\diamond$}}}
\put(1244,129){\raisebox{-.8pt}{\makebox(0,0){$\diamond$}}}
\put(1268,123){\raisebox{-.8pt}{\makebox(0,0){$\diamond$}}}
\put(1293,127){\raisebox{-.8pt}{\makebox(0,0){$\diamond$}}}
\put(1317,123){\raisebox{-.8pt}{\makebox(0,0){$\diamond$}}}
\put(1342,126){\raisebox{-.8pt}{\makebox(0,0){$\diamond$}}}
\put(1366,123){\raisebox{-.8pt}{\makebox(0,0){$\diamond$}}}
\put(1390,125){\raisebox{-.8pt}{\makebox(0,0){$\diamond$}}}
\put(1415,124){\raisebox{-.8pt}{\makebox(0,0){$\diamond$}}}
\put(1439,123){\raisebox{-.8pt}{\makebox(0,0){$\diamond$}}}
\end{picture}
\end{center}
\caption{Mass correction of the first kink mode versus the vacuum modes}
\label{Sine_mode1_contribution}
\end{figure}
The discrete mode i.e. zero mode correction of the Sine-Gordon kink is 
\begin{equation}
\delta m_0=-0.28402
\end{equation}
which is very close to the exact value (\ref{sine_lowest}). The numerical
value of the mass correction is 
\begin{equation}
\delta m~=~-0.318144
\end{equation}
compared to the exact value of -0.318309 and is 99.95\% accurate. This is 
a very satisfactory result. 

\section{Conclusion}

We have shown that the trace formula works very well in (1+1) dimensions for
the Sine-Gordon and the $\phi^4$ kink model. The numerical quantum mass
correction is very close to the exact one. Our technique can be applied with
ease to any (1+1) dimensional theory. This allows us to calculate the mass
correction to {\it non-integrable} solitonic systems, for
example. Specifically, we are interested in the Sine-Skyrme model
\cite{sine:skyrme} and plan to study the mass correction numerically. Or, we
can look at the mass correction of multi-solitons, for example. 

There are two drawbacks. We have used a brute force matrix diagonalisation to
find the eigenvalues. It works very well and is reasonably fast in (1+1)
dimensions. However, if you are to implement the trace formula in higher
dimensions, you will have to use a different technique. Barnes and Turok in
\cite{bt} have used a diffusion  equation method; as discussed in the chapter
on numerical methods. They compute the zero mode, then project it out of the
initial configuration, get the next mode and so on. Only the first modes are
accurate, because the errors are summing up. Computational restrictions also 
limit the calculations to the first few modes. 
 
\bibliographystyle{plain}  
\bibliography{references}
\end{document}